\documentclass[fleqn,twoside]{article}
\usepackage{espcrc2}

\usepackage{graphicx}
\usepackage[figuresright]{rotating}

\newcommand{\AmS}{{\protect\the\textfont2
  A\kern-.1667em\lower.5ex\hbox{M}\kern-.125emS}}

\title{Precision Tests of Flavor and CP Violation in $B$ Decays
\,\thanks{Invited talk presented at the First Workshop on Theory, Phenomenology and
Experiments in Heavy Flavor Physics, Anacapri, Italy, May 29-31, 2006.}}

\author{Michael Gronau\,
\address{\,Physics Department, Technion -- Israel
Institute of Technology, \\
        32000 Haifa, Israel}}%

\begin{document}

\def \app{D_{\pi \pi}}
\def \b{{\cal B}}
\def \bbpp{\overline{{\cal B}}_{\pi \pi}}
\def \bea{\begin{eqnarray}}
\def \beq{\begin{equation}}
\def \bg{\bar \Gamma}
\def \bl{\bar \lambda}
\def \bo{B^0}
\def \ko{K^0}
\def \ob{\overline{B}^0}
\def \lesssim{\stackrel{<}{\sim}}
\def \largesim{\stackrel{>}{\sim}}
\def \bpb{\stackrel{(-)}{B^0}}
\def \cn{Collaboration}
\def \cpp{C_{+-}}
\def \eea{\end{eqnarray}}
\def \eeq{\end{equation}}
\def \ite{{\it et al.}}
\def \kpb{\stackrel{(-)}{K^0}}
\def \lpp{\lambda_{\pi \pi}}
\def \ob{\overline{B}^0}
\def \ok{\overline{K}^0}
\def \rpp{R_{\pi \pi}}
\def \rt{r_{\tau}}
\def \s{\sqrt{2}}
\def \half{\frac{1}{2}}
\def \3half{\frac{3}{2}}
\def \spp{S_{+-}}

\begin{abstract}
Isospin and flavor SU(3) set stringent bounds on penguin pollution in $B^0(t)\to \rho^+\rho^-$,
 providing a theoretically precise determination of $\alpha\equiv \phi_2 $,
 $\alpha = (91\pm 7_{\rm exp}\pm 3_{\rm th})^\circ$. Isospin breaking in a sum rule for
 $B\to K\pi$ rates is shown to be suppressed. A similar sum rule holds for CP asymmetries
 in $B\to K\pi$. Violation of these sum rules would be evidence for an anomalous $\Delta I=1$
 piece in ${\cal H}_{\rm eff}$.

\vspace{1pc}
\end{abstract}

\maketitle

\section{INTRODUCTION}

Two major purposes of high statistics experiments studying $B$ and $B_s$ decays in $e^+e^-$
and hadron colliders are: (1) Achieving great precision in Cabibbo-Kobayashi-Maskawa
parameters~\cite{CKM}, in particular determining the CP violating phase, the source of CP
violation in the Standard Model. (2) Identifying potential
inconsistencies by over-constraining these parameters.
For instance, the phase $\beta\equiv \phi_1 \equiv {\rm arg}(-V^*_{cb}V_{cd}/V^*_{tb}V_{td})
=(21.7^{+1.3}_{-1.2})^\circ$, measured very precisely in time-dependent CP asymmetries of $B^0$
decays via $b \to c \bar cs$~\cite{BS,HFAG}, may be tested also in $b \to sq\bar q~(q=u,d,s)$ penguin-dominated decays~\cite{PenLP,PenMG} which are susceptible to effects of physics beyond the
Standard Model~\cite{GroWo}. Alternatively, a violation of precise symmetry relations among
certain observables in $b\to sq\bar q$ transitions could provide unambiguous
evidence for new physics.

In this talk I wish to discuss a few examples for achieving these
two goals, where important progress has been made recently, both
theoretically and experimentally. In Section 2 I focus on the
currently most precise determination of $\alpha\equiv \phi_2 \equiv
{\rm arg}(-V^*_{tb}V_{td}/V^*_{ub}V_{ud})$, based primarily on $B\to
\rho^+\rho^-$. While the traditional method is based on isospin
symmetry~\cite{GL}, I will argue for an advantage of using flavor
SU(3) to set bounds on the penguin amplitude contributing to this
process~\cite{BGRS}. In Section 3 I study two precision isospin sum
rules, for $B\to K\pi$ decay rates~\cite{GR,Lip} and for the CP
asymmetries in these decays~\cite{GRAsym,KpiAsym}. Isospin breaking
corrections to the sum rule for rates will be shown to be suppressed
by the small ratio of tree and penguin amplitudes contributing in
these decays~\cite{GGRR}.  The sum rule involving four CP
asymmetries~\cite{KpiAsym} (or three asymmetries~\cite{GRAsym})
replaces a much less accurate relation, $A_{\rm CP}(K^+\pi^0) \sim
A_{\rm CP}(K^+\pi^-)$~\cite{GR}, which is sometimes being claimed to
hold in the Standard Model but seems to be violated experimentally.

\section{PRECISION DETERMINATION OF $\alpha$}

\subsection{$B\to\pi\pi$}

The amplitude for $B^0\to\pi^+\pi^-$ contains two terms~\cite{PenLP,PenMG},
conventionally denoted ``tree" ($T$) and ``penguin" ($P$) amplitudes,
involving a weak phase $\gamma$ and a strong phase $\delta$:
\beq\label{Apipi}
A(\bo \to \pi^+ \pi^-) =  T e^{i \gamma} + P e^{i \delta}.
\eeq
We use the $c$-convention~\cite{pipi-con}, in which the top-quark in the $\bar b\to\bar d$
loop has been integrated out and the unitarity relation $V^*_{tb}V_{td} = -V^*_{cb}V_{cd}
-V^*_{ub}V_{ud}$ has been used.
A rather large value, $P/T\sim 0.5$, is implied~\cite{GRalpha} by comparing within
flavor SU(3) the measured rate for this process with that measured for $B\to K\pi$.

Time-dependent decay rates, for an initial $B^0$ or a $\ob$, are
given by~\cite{PenMG}
\bea\label{SC}
&&\Gamma(B^0(t)/\ob(t)\to\pi^+\pi^-) \propto
e^{-\Gamma t}\Gamma_{\pi^+\pi^-}\times\nonumber\\
&&
\left [ 1\pm \cpp\cos\Delta m t \mp \spp\sin\Delta m t\right ],
\eea
\bea
\spp&= &\frac{2 {\rm Im}(\lpp)}{1 + |\lpp| ^2},~~~
\cpp = \frac{1 - |\lpp|^2}{1 + |\lpp|^2},\\
\lpp & \equiv &  e^{-2i \beta} \frac{A(\ob \to \pi^+ \pi^-)}
{A(B^0 \to \pi^+ \pi^-)}.
\eea
The three measurables, $\Gamma_{\pi^+\pi^-}$, $\spp$ and $\cpp$ are
insufficient for determining $T, P, \delta$ and $\gamma$ or $\alpha$.

The isospin method~\cite{GL} is based on obtaining additional information from
two isospin  triangle relations for for $B$ and $\bar B$,
\beq\label{isotr}
A(\pi^+\pi^-)/\s + A(\pi^0\pi^0)-A(\pi^+\pi^0)=0.
\eeq
Defining $\sin 2\alpha_{\rm eff} \equiv
S_{+-}/(1 - C^2_{+-})^{1/2}$, the difference $\theta \equiv \alpha_{\rm eff}-\alpha$
is determined up to a sign ambiguity by constructing the two isospin triangles
with a common base, $A(\pi^+\pi^0)=A(\pi^-\pi^0)$.
A small electroweak penguin amplitude creates a calculable angle between the
two basis, implying a calculable correction in the extracted value of $\alpha$~\cite{EWP},
 $\Delta\alpha_{\rm EWP}=-1.5^\circ$.

In the absence of separate branching ratio measurements for $B^0\to\pi^0\pi^0$ and
$\ob\to\pi^0\pi^0$, the strongest upper bound on $|\theta|$ in terms of CP-averaged rates
and a direct CP asymmetry in $B^0\to\pi^+\pi^-$ is given by~\cite{GLSS}
\beq
\cos 2\theta \ge \frac{\left( {1\over 2}\Gamma_{+-} +
\Gamma_{+0} - \Gamma_{00} \right)^2 -
 \Gamma_{+-}\Gamma_{+0}}{\Gamma_{+-} \Gamma_{+0} \sqrt{1-C^2_{+-}}}~.
\eeq
Somewhat weaker bounds contained in this bound were derived in Ref.~\cite{GQ}.
A complete isospin analysis requires measuring also $C_{00}\equiv -A_{CP}(\pi^0\pi^0)$,
the direct asymmetry in $B^0\to\pi^0\pi^0$.

Current asymmetry measurements~\cite{HFAG}, $\spp=0.50\pm 0.12$,
$\cpp=-0.37\pm 0.10$,
and corresponding branching ratio measurements, imply $\alpha_{\rm eff} = (106\pm
5)^\circ,~|\theta| < 36^\circ$. Two properties, $P/T \le1,~|\delta| \le \pi/2$,  confirmed
experimentally in a global parameter fit, have been shown to resolve a sign ambiguity
in $\theta$~\cite{GLW}, thereby implying $\alpha < \alpha_{\rm eff}$ and consequently
\beq\label{alpha-pipi}
\alpha = (88\pm 18)^\circ.
\eeq

\subsection{Isospin in $B\to\rho\rho$}

Angular analyses of the pions in $\rho$ decays have shown that
$B^0\to \rho^+\rho^-$ is dominated by longitudinal polarization~\cite{HFAG},
$f_L = 0.97^{+0.02}_{-0.03}$. This simplifies the study of
CP asymmetries in these decays (an example of $B\to VV$) to becoming similar to
$B^0\to\pi^+\pi^-$ (an example of $B\to PP$).
The advantage of $B\to \rho\rho$ over $B\to \pi\pi$ is  the smallness of $\b(\rho^0\rho^0)$
($\b(\rho^0\rho^0) <1.1\times 10^{-6}$) relative to $\b(\rho^+\rho^-)$ and $\b(\rho^+\rho^0)$
[both in the range $(20-30)\times 10^{-6}$] in comparison with the corresponding
relative branching ratios in $B\to \pi\pi$. The smaller $P/T$ ratio in $B\to\rho^+\rho^-$,
$P/T< 0.3$~\cite{CKMfit}, leads to a stronger upper bound $|\alpha-\alpha_{\rm eff}|< 11^\circ$
in $B\to\rho\rho$. The asymmetry measurements~\cite{HFAG},
$S_L = -0.21 \pm 0.22,~C_L = -0.03 \pm 0.17$, lead to
$\alpha_{\rm eff} = (96^{+7}_{-6})^\circ$. Thus, one finds $\alpha = (96 \pm 13)^\circ$ by adding
errors in quadrature. This error is dominated by the current uncertainty in $\alpha-\alpha_{\rm eff}$
including an ambiguity in its sign.

The error in $\alpha$ may be reduced by making one very reasonable and mild assumption about
the strong phase, $|\delta|\le \pi/2$. This is motivated by QCD factorization~\cite{BBNS} where
$\delta$ is suppressed by $1/m_b$ or by $\alpha_s$, and is found to hold in a global SU(3)
fit to $B\to PP$~\cite{SU3PP}. This expected property and $P/T < 1$ imply~\cite{GLW} $\alpha
< \alpha_{\rm eff}$ and consequently
\beq\label{alpha-iso}
\alpha = [90^{+7}_{-6}({\rm exp}) \pm 6({\rm th})]^\circ.
\eeq

\subsection{Bounds on $P/T$ from $B\to K^{*0}\rho^+$}

It has been recently noted~\cite{BGRS} that a stronger constraint on
$P/T$ in $B^0\to\rho^+\rho^-$ may be obtained by relating this
process to $B^+\to K^{*0}\rho^+$ within flavor SU(3). The advantage
of using this process over using $B^0\to\rho^0\rho^0$ is twofold.
First, penguin amplitudes in $\Delta S=1$ decays are enhanced by a
factor $V_{cs}/V_{cd}$ relative to $\Delta S=0$ decays. Second,
$B^+\to K^{*0}\rho^+$ is expected to be dominated by a penguin
amplitude~\cite{GHLR}, whereas in $B^0\to \rho^0\rho^0$ a penguin
amplitude interferes with a potentially comparable or larger
color-suppressed tree amplitude.

One uses both the branching ratio measured for this
process~\cite{HFAG}, $\b(K^{*0}\rho^+)= (9.3 \pm 1.7)\times
10^{-6}$, and the measured fraction of longitudinal rate,
$f_L(K^{*0}\rho^+)= 0.48^{+0.09}_{-0.08}$, to define a CP-averaged
longitudinal amplitude $A_L(K^{*0}\rho^+)$. Since this strangeness
changing process is dominated by a penguin amplitude which is
related by SU(3) to $P$ in $B^0\to\rho^+\rho^-$, a parameter $F$ can
be introduced defined by
\beq
|A_L(K^{*0}\rho^+)|^2 = F\left (\frac{|V_{cs}|f_{K^*}}{|V_{cd}|f_{\rho}}\right )^2P^2.
\eeq
For a given value of $F$, supplementing $\Gamma_L(\rho^+\rho^-), S_L(\rho^+\rho^-),
 C_L(\rho^+\rho^-)$ by $\Gamma_L(K^{*0}\rho^+)$, permits a determination of $\alpha$,
up to a discrete abmiguity~\cite{BGRS}.  The uncertainty in $F$ is the source for a
theoretical error in $\alpha$.

The parameter $F$ equals one in the limit of a purely factorized penguin amplitude.
The ratios $|V_{cs}|/|V_{cd}|$ and $f_{K^*}/f_{\rho}$ describe the corresponding
CKM factors and an SU(3) breaking factor. Corrections to $F=1$ are expected to be small.
They follow from non-factorized SU(3) breaking (or form factor effects) and from two
small terms~\cite{GHLR}, a color-suppressed electroweak penguin amplitude ($P^c_{EW}$)
and a penguin annihilation contribution ($PA$) which is formally $1/m_b$-suppressed.
[Such a contribution would show-up in longitudinally polarized $B_s\to\rho^+\rho^-$
decays~\cite{GHLR}.]
A random scan through the input parameter space describing a model for these contributions in
a QCD-factorization calculation~\cite{F-QCDF} permits a rather broad range, $0.3 \le F \le 1.5$,
favoring values smaller than one over values larger than one. We shall allow an even broader
range which is symmetric around $F=1$,
\beq\label{Frange}
0.3 \le F \le 3.0.
\eeq
This range, which we consider very conservative, will be used to demonstrate the low sensitivity
of the error in $\alpha$ to the uncertainty in $F$.

To appreciate the advantage of this method over the isospin method, we note
that the measurement of $\Gamma_L(K^{*0}\rho^+)$ implies a smaller value for $P/T$ than
implied by $\b(\rho^0\rho^0)$.
A value $F=1$ corresponds to $P/T=0.09$, which is considerably smaller than the upper bound
$P/T<0.3$ obtained from $\b(\rho^0\rho^0)$. Smaller values of $F$ imply a somewhat larger
$P/T$. However, a value $P/T=0.3$ would require $F=(0.09/0.3)^2=0.09$ which is unreasonably
small. The main point here is therefore the following. {\em Once $P$ has been established to be small,
a large relative uncertainty in this amplitude, obtained by assuming flavor SU(3) and neglecting
smaller terms, leads to only a small uncertainty in $\alpha$}.
This expectation is demonstrated by fitting $\Gamma_L(\rho^+\rho^-), S_L(\rho^+\rho^-),
C_L(\rho^+\rho^-)$ and $\Gamma_L(K^{*0}\rho^+)$ in terms of $T, P, \delta$ and $\alpha$
while varying $F$ in the range (\ref{Frange}). The final result is
\beq\label{alpha-SU3}
\alpha = [91^{+9}_{-7}({\rm exp})^{+2}_{-4}({\rm th})]^\circ,
\eeq
where the theoretical error corresponds to the range (\ref{Frange}).
A discrete ambiguity between two values of $\alpha$ corresponding to $|\delta|\le\pi/2$ and
$|\delta|>\pi/2$ has been resolved by assuming $|\delta|\le\pi/2$, as has already been assumed
when applying the isospin method.

\subsection{Averaged value of $\alpha$ from $B^0\to\rho^+\rho^-$}

The two ways described in the previous subsections for extracting $\alpha$ in
$B^0(t)\to \rho^+\rho^-$ provide two independent  constraints on the effect of
the penguin amplitude on the asymmetries $S_L(\rho^+\rho^-)$ and $C_L(\rho^+\rho^-)$.
Taking the average of (\ref{alpha-iso}) and (\ref{alpha-SU3}) one has
\beq\label{alpha}
\alpha = [91 \pm 7({\rm exp}) \pm 3({\rm th})]^\circ.
\eeq

This value assumes that $\delta$ lies in the positive
semicircle, $|\delta|<\pi/2$, an assumption motivated by QCD factorization and by
a global SU(3) fit to $B\to PP$. This weak assumption is expected to be relaxed
further by improving the precision of $C_L(\rho^+\rho^-)$, which would eventually
require excluding only values of $\delta$ near $\pi$. The assumption $|\delta|<\pi/2$
can be tested by fitting within SU(3) decay rates and CP asymmetries for all
$B^{+,0}\to \rho\rho, K^*\rho$ decays for longitudinally polarized vector mesons.

The value (\ref{alpha}) is consistent with values obtained in two global fits to all
other CKM constraints~\cite{CKMfit,UTfit} including the recent measurement of
$\Delta m_s$~\cite{Dms}. Our extracted value is more precise than a value,
$\alpha =(100^{+15}_{-9})^\circ$~\cite{CKMfit}, obtained in a fit combining
$B\to \rho\rho, \pi\pi, \rho\pi$. The result (\ref{alpha}) lies on the
low side of this range because it assumes $|\delta|<\pi/2$. Including $B\to\pi\pi,
\rho\pi$ would affect the
average (\ref{alpha}) only slightly since the error in (\ref{alpha-pipi}) and errors
of similar magnitudes involved in studies of $B\to \rho\pi$~\cite{QS,GZrp} are
considerably larger than the errors in (\ref{alpha-iso}) and (\ref{alpha-SU3}).

The rather small theoretical error in $\alpha$, $\pm 3^\circ$, implies that in
studies using isospin symmetry one must
consider isospin breaking effects which are expected to be of similar magnitude.
As mentioned, in $B\to\pi\pi$ and $B\to\rho\rho$ the effect of electroweak penguin
amplitudes on the isospin analyses has been calculated and was found
to be $\Delta\alpha_{\rm EWP}=-1.5^\circ$. Other effects include $\pi^0$-$\eta$-$\eta'$
mixing which affects the isospin analysis of $B\to\pi\pi$~\cite{Gardner} (a correction
smaller than $1^\circ$ was calculated in~\cite{GZI}), $\rho$-$\omega$ mixing
affecting the $\pi^+\pi^-$ invariant-mass distribution in $B^+\to \rho^+\rho^0$~\cite{GZI},
and a correction to the isospin analysis of $B\to \rho\rho$ from a
potential $I=1$ final state when the two $\rho$ mesons are observed with different
invariant-masses~\cite{FLNQ}. We note that the extraction of $\alpha$ by applying flavor
SU(3) to $B^0\to\rho^+\rho^-$ and $B^+\to K^{*0}\rho^+$ involves only charged $\rho$
mesons, and is therefore not susceptible to correction of this kind.

\section{PRECISION $B\to K\pi$ SUM RULES}

The decays $B\to K\pi$, which are dominated by a $\bar b\to\bar s q\bar q$ penguin
amplitude, are sensitive to physics beyond the Standard Model because new heavy
particles may replace the $W$ boson and the top quark in the loop. In the Standard
Model isospin symmetry implies relations among amplitudes, among rates and among
CP asymmetries in $B\to K\pi$ decays. Relations of this kind are very useful
as their violation would provide evidence for new physics. Sum rule for rates test the flavor
structure of the effective weak Hamiltonian. Symmetry relations among CP
asymmetries are particularly interesting because almost any extension of the model involves
new sources of CP violation which lead to potential deviations  from the sum rules.

\subsection{Isospin in $B\to K\pi$}

The four physical $B\to K\pi$ decay amplitudes are expressed in terms of three
isospin-invartiant amplitudes~\cite{Kpi-iso},
\bea
-A(K^+\pi^-) &=& B_{1/2} - A_{1/2} - A_{3/2},\nonumber\\
A(K^0\pi^+) &=& B_{1/2} + A_{1/2} - A_{3/2},\nonumber\\
-\sqrt{2}A(K^+\pi^0) &=& B_{1/2} + A_{1/2} - 2A_{3/2},\nonumber\\
\sqrt{2}A(K^0\pi^0) &=& B_{1/2} - A_{1/2} + 2A_{3/2},
\eea
where $B, A$ correspond to $\Delta I=0, 1$ parts of ${\cal H}_{\rm eff}$,
respectively, while subscripts denote the isospin of the final $K\pi$ state.
This implies a quadrangle relation,
\bea\label{SigmaKpi}
&& \Sigma A(K\pi)\equiv A(K^+\pi^-) - A(K^0\pi^+)  \nonumber\\
&& -\sqrt{2}A(K^+\pi^0) +\sqrt{2}A(K^0\pi^0) = 0.
\eea
{\em This relation holds separately for $B$ and $\bar B$ decays for any
linear combination of two arbitrary $\Delta I=0$ and $\Delta I=1$
transition operators}. This feature turns out to be crucial when discussing isospin
breaking effects in $B\to K\pi$ sum rules.

\subsection{Precise sum rule for $B\to K\pi$ rates}

The following relation is obeyed approximately for $B\to K\pi$ decay rates~\cite{GR,Lip},
\bea\label{SR}
&& \Gamma(K^+\pi^-) + \Gamma(K^0\pi^+) \approx\nonumber\\
&&  2\Gamma(K^+\pi^0)+2\Gamma(K^0\pi^0).
\eea
This sum rule holds up
to terms which are quadratic in small quantities. The proof of this
sum rule is rather simple. The dominant isospin amplitude is the
singlet $B_{1/2}$, the only one containing a penguin amplitude.
Terms which are linear or quadratic in $B_{1/2}$ involve the product
of this amplitude with the sum $\Sigma A(K\pi)$ which vanishes. The
remaining terms are quadratic in two small ratios of tree ($T$) or
electroweak penguin amplitudes ($P_{EW}$) and the dominant penguin
amplitude ($P$). The two ratios, $T/P$ and $P_{EW}/P$, are between
0.1 and 0.2. Thus, small corrections to the sum rule were calculated
and were found to be between one and five
percent~\cite{F-QCDF,GRKpi,BRS,Williamson:2006hb}.
A larger deviation would require an anomalously large $\Delta I=1$
operator in the effective Hamiltonian.

At the level of precision expected in the Standard Model one must
consider also isospin breaking corrections to the sum rule. In
general one would expect these corrections to be linear in isospin
breaking, namely of order $(m_d-m_u)/\Lambda_{\rm QCD} \simeq 0.03$.
As it turns out, isospin breaking corrections in the sum rule are
further suppressed by the small ratio $T/P$. The argument for this
suppression holds for both Eqs.~(\ref{SigmaKpi}) and
(\ref{SR})~\cite{GGRR} and is explained briefly in the next two
paragraphs.

The spurion representing isospin breaking caused by the mass and charge-difference
of the $d$ and $u$ quarks behaves like a sum of $\Delta I=0$ and $\Delta I=1$
operators. Thus, in the presence of isospin breaking the dominant isosinglet amplitude
$B_{1/2}$ becomes a sum of $\Delta I =0$ and $\Delta I =1$ amplitudes. Since
Eq.~(\ref{SigmaKpi}) holds for an arbitrary combination of isosinglet and isotriplet
operators, it holds for the dominant terms also when isospin breaking is included.
Thus, isospin breaking appears only in subdominant terms in (\ref{SigmaKpi}) and
is suppressed by $T/P$.

Similarly, the dominant terms in Eq.~(\ref{SR}) are quadratic in $B_{1/2}$ and their
linear isospin breaking term is a combination of isosinglet and isotriplet
contributions. The sum of contributions appearing in (\ref{SR}) vanishes as it
involves the same combination as (\ref{SigmaKpi}). Since isospin breaking corrections
cancel in (\ref{SR}) in terms which are quadratic in $B_{1/2}$, the remaining isospin
breaking is suppressed by $T/P$ (or by $A_{1/2,3/2}/B_{1/2}$). This correction is
significantly smaller than that of the quadratic terms correcting the sum rule which are
between one and five percent.

The current experimental situation of the sum rule (\ref{SR}) can be summarized in
terms of branching ratios corrected by the $B^+/B^0$ lifetime
ratio, $\tau_+/\tau_0=1.076\pm 0.008$~\cite{HFAG},
\beq
44.4 \pm 1.5 \approx 48.9\pm 2.7,
\eeq
which works within $1.5\sigma$. Two kinds of isospin breaking effects  in branching
ratio measurements must be studied more carefully: (i) Radiative
corrections~\cite{BI} which have not been included in all $B\to K\pi$ measurements.
(ii) The effect on branching ratio measurements of a small isospin-breaking difference
between the production rates of $B^+B^-$ and $B^0\ob$ pairs
at the $\Upsilon(4S)$~\cite{GGR}.

\subsection{Success of SU(3) in CP asymmetries}

CP asymmetries in $B\to K\pi$ decays have been the subject of a large number of theoretical
studies, whose most difficult parts were model-dependent calculations of strong phases.
While it is hard to calculate magnitudes of asymmetries, one may
compare several asymmetries using symmetry arguments.
Of the four $B\to K\pi$ asymmetries only one, that measured in $B^0\to K^+\pi^-$
is significantly different from zero~\cite{HFAG}, $A_{\rm CP}(K^+\pi^-) = -0.108\pm 0.017$.
It is interesting to compare this asymmetry with a second nonzero asymmetry measured in $B^0\to\pi^+\pi^-$, $A_{\rm CP}(\pi^+\pi^-)=0.37\pm 0.10$. (The error does not reflect
a certain disagreement between Babar and Belle measuremnts.)
In flavor SU(3),  the two processes involve common tree and penguin amplitudes multiplied by
different CKM facors.
While the amplitude of $B^0\to\pi^+\pi^-$ is given by (\ref{Apipi}), that of $B^0\to K^+\pi^-$ is
\beq
A(B^0\to K^+\pi^-) = \lambda Te^{i\gamma} - \lambda^{-1}Pe^{i\delta},
\eeq
where $\lambda = V_{us}/V_{ud}$. Consequently, the two CP rate differences have equal
magnitudes and opposite signs,  and the asymmetries are related
in a reciprocal manner to the corresponding branching ratios~\cite{GHLR,DH},
\bea\label{SU3Asym}
\frac{A_{\rm CP}(\pi^+\pi^-)}{A_{\rm CP}(K^+\pi^-)} &=& -\frac{\b(K^+\pi^-)}{\b(\pi^+\pi^-)},
\nonumber\\
-3.4\pm 1.1 &=& -4.0\pm 0.4.
\eea
The agreement of signs and magnitudes supports the assumption that weak hadronic
amplitudes and strong phases are approximately SU(3) invariant. Deviations at a level
of $30\%$ are expected in (\ref{SU3Asym}) from SU(3) breaking corrections and from
$1/m_b$-suppressed annihilation amplitudes.

\subsection{Precise sum rule for $K\pi$ asymmetries}

Isospin symmetry can be applied to the four $K\pi$ asymmetries to
obtain an approximate relation~\cite{GRAsym,KpiAsym,AS},
\beq\label{SRA} 2(\Delta_{+0}+\Delta_{00})\approx \Delta_{+-} +
\Delta_{0+},
\eeq
where one defines CP rate differences
$\Delta_{ij}\equiv \Gamma(B\to K^i\pi^j) - \Gamma(\bar B\to K^{\bar
i}\pi^{\bar j})$. The proof of this sum rule is based on the fact
that each CP rate difference can be written as a product of an
imaginary part of products of hadronic amplitudes and an invariant
imaginary part of products of CKM factors, $4{\rm
Im}(V^*_{tb}V_{ts}V_{ub}V^*_{us})$. Writing the dominant term in
$\Delta_{ij}$ symbolically as ${\rm IM}[P^*A(K^i\pi^j)]$, where $P$
dominates each of the $K\pi$ amplitudes, the dominant term in the
difference between the left-hand side and the right-hand-side of
(\ref{SRA}) is ${\rm IM}[P^*\Sigma A(K\pi)]$ which vanishes by
(\ref{SigmaKpi}). The subdominant terms in this difference,
involving terms of the form ${\rm IM}(P^*_{EW}T)$, are suppressed by
about an order of magnitude relative to ${\rm IM}(P^*T)$ and can be
shown to cancel in the flavor SU(3) limit.

In the penguin dominance approximation, $\Gamma(K^+\pi^-)\approx\Gamma(K^0\pi^+)
\approx 2\Gamma(K^+\pi^0)\approx 2\Gamma(K^0\pi^0)$,
the relation (\ref{SRA}) simplifies to a sum rule among corresponding CP asymmetries,
\bea
&& A_{\rm CP}(K^+\pi^0) + A_{\rm CP}(K^0\pi^0) \approx\nonumber\\
&&  A_{\rm CP}(K^+\pi^-) + A_{\rm CP}(K^0\pi^+).
\eea
This sum rule is expected to hold to any foreseeable experimental precision, the
most difficult asymmetry being that of $B^0\to K^0\pi^0$. Using the currently
measured asymmetries~\cite{HFAG}, it reads
\bea
&& (0.04\pm 0.04) + (0.02\pm 0.13) \approx \nonumber\\
&& (-0.108\pm 0.017) + (-0.02\pm 0.04),
\eea
which holds within experimental errors. Using three of the measured asymmetries, the
sum rule (\ref{SRA}) predicts $A_{\rm CP}(K^0\pi^0) = -0.15\pm 0.06$.

Before closing this section we wish to comment briefly on a
so-called ``puzzle" which is sometimes being claimed by observing
$A_{\rm CP}(K^+\pi^0)\ne A_{\rm CP}(K^+\pi^-)$. The approximation
$A_{\rm CP}(K^+\pi^0) \sim A_{\rm CP}(K^+\pi^-)$ was suggested
several years ago~\cite{GR} based on classifying contributions to
$B\to K\pi$ amplitudes in terms of flavor topologies~\cite{GHLR},
where a hierarchy $C\ll T$ was assumed between color-allowed and
color-suppressed tree amplitudes. In fact, there exists no
compelling theoretical argument for a suppression of $C$ relative to
$T$. A global SU(3) analysis of $B\to PP$~\cite{SU3PP,BHLDS}
indicates that the two contributions are comparable. An example for
a sizable $C$~\cite{BFRS} is the large $\b(\pi^0\pi^0)$. Abandoning
the assumption $C\ll T$, the two asymmetries $A_{\rm CP}(K^+\pi^0)$
and $A_{\rm CP}(K^+\pi^-)$ could be different in the Standard Model.
The sum rule (\ref{SRA}) (or a similar sum rule in which a small
$\Delta_{0+}$ is neglected~\cite{GRAsym}) holds also for a sizable
$C$.

 \section{CONCLUSION}

 The currently most precise extraction of $\alpha$  based on $B^0\to\rho^+\rho^-$
 involves a theoretical error of a few degrees ($\pm 3^\circ$), requiring the inclusion
 of isospin breaking
 corrections where applicable. Two sum rules were studied based on isospin symmetry, involving
 $B\to K\pi$ decay rates and CP asymmetries in these processes.
Isospin breaking in the first sum rule was shown to be suppressed and is therefore
 negligible, while the second sum rule is expected to hold within any foreseeable experimental
 accuracy.

 Both sum rules are unaffected by a new isoscalar operator, which could
 simply be added to the dominant penguin amplitude. A potential violation of the sum rules
 would therefore be evidence for an anomalous $\Delta I=1$ operator in the
 effective Hamiltonian. Observing such a violation requires reducing experimental
 errors in $B\to K\pi$ rates and asymmetries by at least a factor two.

 \medskip
I wish to thank the SLAC Theory Group for its kind hospitality. This
work was supported in part by the Israel Science Foundation under
Grant  No. 1052/04 and by the German-Israeli Foundation under Grant
No. I-781-55.14/2003.

\def \ajp#1#2#3{Am.\ J. Phys.\ {\bf#1} (#3) #2}
\def \apny#1#2#3{Ann.\ Phys.\ (N.Y.) {\bf#1}, #2 (#3)}
\def \app#1#2#3{Acta Phys.\ Polonica {\bf#1}, #2 (#3)}
\def \arnps#1#2#3{Ann.\ Rev.\ Nucl.\ Part.\ Sci.\ {\bf#1}, #2 (#3)}
\def \art{and references therein}
\def \cmts#1#2#3{Comments on Nucl.\ Part.\ Phys.\ {\bf#1}, #2 (#3)}
\def \cn{Collaboration}
\def \econf#1#2#3{Electronic Conference Proceedings {\bf#1}, #2 (#3)}
\def \efi{Enrico Fermi Institute Report No.}
\def \epjc#1#2#3{Eur.\ Phys.\ J.\ C {\bf#1} (#3) #2}
\def \ib{{\it ibid.}~}
\def \ibj#1#2#3{~{\bf#1}, #2 (#3)}
\def \ijmpa#1#2#3{Int.\ J.\ Mod.\ Phys.\ A {\bf#1} (#3) #2}
\def \ite{{\it et al.}}
\def \jhep#1#2#3{JHEP {\bf#1} (#3) #2}
\def \jpb#1#2#3{J.\ Phys.\ B {\bf#1}, #2 (#3)}
\def \mpla#1#2#3{Mod.\ Phys.\ Lett.\ A {\bf#1} (#3) #2}
\def \nat#1#2#3{Nature {\bf#1}, #2 (#3)}
\def \nc#1#2#3{Nuovo Cim.\ {\bf#1}, #2 (#3)}
\def \nima#1#2#3{Nucl.\ Instr.\ Meth.\ A {\bf#1}, #2 (#3)}
\def \npb#1#2#3{Nucl.\ Phys.\ B~{\bf#1}  (#3) #2}
\def \npps#1#2#3{Nucl.\ Phys.\ Proc.\ Suppl.\ {\bf#1} (#3) #2}
\def \PDG{Particle Data Group, K. Hagiwara \ite, \prd{66}{010001}{2002}}
\def \pisma#1#2#3#4{Pis'ma Zh.\ Eksp.\ Teor.\ Fiz.\ {\bf#1}, #2 (#3) [JETP
Lett.\ {\bf#1}, #4 (#3)]}
\def \pl#1#2#3{Phys.\ Lett.\ {\bf#1}, #2 (#3)}
\def \pla#1#2#3{Phys.\ Lett.\ A {\bf#1}, #2 (#3)}
\def \plb#1#2#3{Phys.\ Lett.\ B {\bf#1} (#3) #2}
\def \prl#1#2#3{Phys.\ Rev.\ Lett.\ {\bf#1} (#3) #2}
\def \prd#1#2#3{Phys.\ Rev.\ D\ {\bf#1} (#3) #2}
\def \prp#1#2#3{Phys.\ Rep.\ {\bf#1} (#3) #2}
\def \ptp#1#2#3{Prog.\ Theor.\ Phys.\ {\bf#1} (#3) #2}
\def \rmp#1#2#3{Rev.\ Mod.\ Phys.\ {\bf#1} (#3) #2}
\def \rp#1{~~~~~\ldots\ldots{\rm rp~}{#1}~~~~~}
\def \yaf#1#2#3#4{Yad.\ Fiz.\ {\bf#1}, #2 (#3) [Sov.\ J.\ Nucl.\ Phys.\
{\bf #1}, #4 (#3)]}
\def \zhetf#1#2#3#4#5#6{Zh.\ Eksp.\ Teor.\ Fiz.\ {\bf #1}, #2 (#3) [Sov.\
Phys.\ - JETP {\bf #4}, #5 (#6)]}
\def \zp#1#2#3{Zeit.\ Phys.\ {\bf#1} (#3) #2}
\def \zpc#1#2#3{Zeit.\ Phys.\ C {\bf#1} (#3) #2 }
\def \zpd#1#2#3{Zeit.\ Phys.\ D {\bf#1}, #2 (#3)}


\begin{thebibliography}{99}

\bibitem{CKM} N. Cabibbo, \prl{10}{531}{1963};
M. Kobayashi and T. Maskawa, \ptp{49}{652}{1973}.

\bibitem{BS}  A. B. Carter and A. I. Sanda, \prd{23}{1567}{1981};
I. I. Bigi and A. I. Sanda, \npb{193}{85}{1981}.

\bibitem{HFAG} E. Barbiero {\it et al.}, Heavy Flavor Averaging Group,
arXiv:hep-ex/0603003; updated in {\tt www.slac.stanford.edu/xorg/hfag.}

\bibitem{PenLP} D. London and R. D. Peccei, \plb{223}{257}{1989}.

\bibitem{PenMG} M. Gronau, \prl{63}{1451}{1989}.

\bibitem{GroWo} M. Gronau and D. London, \prd{55}{2845}{1997}; Y. Grossman
and M. Worah, \plb{395}{241}{1997}; D. London and A. Soni, \plb{407}{61}{1997}.

\bibitem{GL} M. Gronau and D. London, \prl{65}{3381}{1990}.

\bibitem{BGRS} M. Beneke, M. Gronau, J. Rohrer and M. Spranger, \plb{638}{68}{2006}.

\bibitem{GR} M. Gronau and J. L. Rosner, \prd{59}{113002}{1999}.

\bibitem{Lip}
H. J. Lipkin, \plb{445}{403}{1999}.

\bibitem{GRAsym} M. Gronau and J. L. Rosner, \prd{71}{074019}{2005}.

\bibitem{KpiAsym} M. Gronau, \plb{627}{82}{2005}.

\bibitem{GGRR} M. Gronau, Y. Grossman, G. Raz and J. L. Rosner, \plb{635}{207}{2006}.

\bibitem{pipi-con} M. Gronau and J. L. Rosner, \prd{66}{053003}{2002}.

\bibitem{GRalpha}  M. Gronau and J. L. Rosner, \plb{595}{339}{2004}.

\bibitem{EWP} A. Buras and R. Fleischer, \epjc{11}{93}{1999}; M. Gronau, D. Pirjol
and T. M. Yan, \prd{60}{034021}{1999}.

\bibitem{GLSS}M. Gronau, D. London, N. Sinha and R. Sinha,
\plb{514}{315}{2001}.

\bibitem{GQ}  Y. Grossman and H. R. Quinn, \prd{58}{017504}{1998}; J. Charles, \prd{59}{054007}{1999}.

\bibitem{GLW} M. Gronau, E. Lunghi and D. Wyler, \plb{606}{95}{2005}.

\bibitem{CKMfit} J. Charles \ite, CKMfitter Group,
\epjc{41}{1}{2005}; updated in
{\tt www.slac.stanford.edu/xorg/ckmfitter.}

\bibitem{BBNS} M. Beneke, G. Buchalla, M. Neubert, and C. T. Sachrajda,
\prl{83}{1914}{1999}; \npb{606}{245}{2001}.

\bibitem{SU3PP} C.~W.~Chiang, M.~Gronau, J.~L.~Rosner and D.~A.~Suprun,
\prd{70}{034020}{2004}.

\bibitem{GHLR} M.~Gronau, O.~F.~Hernandez, D.~London and J.~L.~Rosner,
\prd{50}{4529}{1994};  \ib {\bf 52} (1995) 6374.

\bibitem{F-QCDF} M. Beneke and M. Neubert, \npb{675}{333}{2003}.

\bibitem{UTfit} M. Bona {\it et al.}, hep-ph/0606167.

\bibitem{Dms} A. Abulencia {\it et al.}, CDF Collaboration, hep-ex/0606027.

\bibitem{QS} H. R. Quinn and A. E. Snyder, \prd{48}{2139}{1993}.
B. Aubert {\it et al.}, Babar Collaboration, hep-ex/0408099.

\bibitem{GZrp} M. Gronau and J. Zupan, \prd{70}{074031}{2004}.

\bibitem{Gardner} S. Gardner, \prd{59}{077502}{1999}; \prd{72}{034015}{2005}.

\bibitem{GZI} M. Gronau and J. Zupan, \prd{71}{074017}{2005}.

\bibitem{FLNQ} A. F. Falk, Z. Ligeti, Y. Nir and H. R. Quinn,
\prd{69}{011502(R)}{2004}.

\bibitem{Kpi-iso} H. J. Lipkin, Y. Nir, H. R. Quinn and A. Snyder, \prd{44}{1454}{1991};
M. Gronau, \plb{265}{389}{1991}.

\bibitem{GRKpi} M. Gronau and J. L. Rosner, \plb{572}{43}{2003}.

\bibitem{BRS} C. W. Bauer, I. Z. Rothstein and I. W. Stewart, hep-ph/0510241.

\bibitem{Williamson:2006hb}
A.~R.~Williamson and J.~Zupan, \prd{74}{014003}{2006}.

\bibitem{BI} E. Baracchini and G. Isidori, \plb{633}{309}{2006}.

\bibitem{GGR} M. Gronau, Y. Grossman and J. L. Rosner, \prd{73}{057501}{2006}.

\bibitem{DH} N. Deshpande and X. G. He, \prl{75}{1703}{1995}.

\bibitem{AS} D. Atwood and A. Soni, \prd{58}{036005}{1998}.

\bibitem{BHLDS} S. Baek, P. Hamel, D. London, A. Datta and D. Suprun, \prd{71}{057502}{2005}.

\bibitem{BFRS} A. Buras, R. Fleischer, S. Recksiegel and F. Schwab, \prl{92}{101804}{2004};
\npb{697}{133}{2004}.

\end{thebibliography}
\end{document}